\documentclass[]{article}

\usepackage[margin=0.75in]{geometry}
\usepackage[capposition=top]{floatrow}
\usepackage{xcolor}
\usepackage{amsmath}
\usepackage{amsfonts}
\usepackage{amsthm}
\usepackage{graphicx}
\usepackage{float}
\usepackage{hyperref}
\usepackage{setspace}
\usepackage{multicol}
\usepackage{lipsum}
\usepackage{algorithm}
\usepackage{algpseudocode}

\theoremstyle{definition}

\title{Confidence Regions for Multiple Outcomes, Effect Modifiers, and Other Multiple Comparisons}
\author{Paul N Zivich\textsuperscript{1}, Stephen R Cole\textsuperscript{1}, Noah Greifer\textsuperscript{2}, Lina M Montoya\textsuperscript{3,4}, Michael R Kosorok\textsuperscript{4}, \\ Jessie K Edwards\textsuperscript{1}}
\date{%
	\textsuperscript{1}Department of Epidemiology, Gillings School of Global Public Health, University of North Carolina at Chapel Hill, Chapel Hill, NC, USA\\%
	\textsuperscript{2}The Institute for Quantitative Social Science, Harvard University, Cambridge, MA, USA\\%
	\textsuperscript{3}School of Data Science and Society, Chapel Hill, NC, USA\\%
	\textsuperscript{4}Department of Biostatistics, Gillings School of Global Public Health, University of North Carolina at Chapel Hill, Chapel Hill, NC, USA\\%
	~\\
	\today
}

\begin{document}
	
\maketitle

\begin{abstract}
	In epidemiology, some have argued that multiple comparison corrections are not necessary as there is rarely interest in the universal null hypothesis. From a parameter estimation perspective, epidemiologists may still be interested in multiple parameters. In this context, standard confidence intervals are not guaranteed to provide simultaneous coverage of more than one parameter. In other words, use of confidence intervals in these cases will understate the uncertainty due to random error. To address this challenge, one can use confidence bands, an extension of confidence intervals to parameter vectors. We illustrate the use of confidence bands in three case studies: estimation of multiple causal effects, effect measure modification by a binary variable, and effect measure modification by a continuous variable. Each example uses publicly available data is accompanied by SAS, R, and Python code. The type of confidence region reported by epidemiologists should depend on whether scientific interest is in a single parameter or a set of parameters. For sets of parameters, like in cases where multiple actions or outcomes, effect measure modification, dose-response, or other functions are of interest, sup-t confidence bands are preferred due to their statistical properties, computational simplicity, and ease of presentation.
\end{abstract}

\section*{Introduction}

In the inaugural issue of \textit{Epidemiology}, Kenneth Rothman made the argument that multiple comparison corrections are not to be recommended \cite{rothmanNoAdjustmentsAre1990}, which has been a subject of much debate \cite{greenlandEmpiricalBayesAdjustmentsMultiple1991, pooleMultipleComparisonsNo1991, savitzMultipleComparisonsRelated1995, feiseMultipleOutcomeMeasures2002, greenlandMultipleComparisonsAssociation2008, youngCerealinducedGenderSelection2009, youngDemingDataObservational2011, goldbergMultipleComparisonsDesign2011, gelmanWhyWeUsually2012, farlandPvaluesReproductiveHealth2016, althouseAdjustMultipleComparisons2016, sjolanderFrequentistBayesianApproaches2019, greenlandMultipleComparisonsControversies2019, sjolanderAuthorsReplyMultiple2020, hooperAdjustNotAdjust2025}. Here, we consider the case of multiple comparison corrections in the context of a single study, as the analytical decisions for this process are under the control of a single set of investigators. In this context, Rothman argues that the universal null hypothesis undermines the intention behind empirical research due to the inflation of false negative results since it is unlikely for the collection of all relationships under consideration to be null. In light of the movement in epidemiology from null hypothesis significance testing to focusing on parameter estimation to answer specific scientific questions of interest \cite{gardnerConfidenceIntervalsRather1986, stangStatisticalInferenceAbstracts2017b}, we revisit this concern over multiple comparisons.

Rather than testing the universal null hypothesis, multiple comparisons can also be thought of as estimating a set, or vector, of parameters. Common examples of when a vector of parameters is of interest include when there are multiple outcomes of interest \cite{menyhartMultiplicityCorrectionsLife2025}, when there are multiple actions or regimes of interest \cite{montoyaEfficientRobustApproaches2023}, when studying effect measure modification of joint effects \cite{menyhartMultiplicityCorrectionsLife2025, boursTutorialNontechnicalExplanation2021}, when a regression model entails multiple parameters being reported \cite{hothornSimultaneousInferenceGeneral2008}, or when reporting functions (e.g., survival, hazard, dose-response) \cite{sachsConfidenceBandsSurvival2022, ashConfidenceBandsHypothesis2022, lieblFastFairSimultaneous2023, caiOnestepTargetedMaximum2020a}. In each of these cases, parameter estimates are often accompanied by confidence intervals for inference. Under a frequentist perspective, if our study were repeated indefinitely then $(1-\alpha) \times 100\%$ of the $(1-\alpha) \times 100\%$ confidence intervals would contain the true parameter absent systematic errors \cite{sedgwickUnderstandingConfidenceIntervals2014}. However, things change when considering parameter vectors. For parameter vectors, confidence intervals for a single parameter only have nominal coverage for \textit{that} parameter. The collection of confidence intervals does not provide nominal \textit{simultaneous} coverage for \textit{all} parameters in the vector. While we agree with Rothman that the universal null (testing whether \textit{all} parameters in the vector are equal to zero) might not be useful in epidemiology, we see reason to be concerned when reporting multiple parameters with confidence intervals that no longer provide correct simultaneous coverage. Explicitly, our reported results would be understating the uncertainty due to random error. Others have made similar objections to Rothman's argument in the context of parameter estimation \cite{greenlandEmpiricalBayesAdjustmentsMultiple1991}. While we know that these confidence intervals are too narrow, how much wider the intervals should be is context dependent.

Here, we provide an introduction to confidence regions intended to adjust our uncertainty about parameter vectors with respect to random error. The reviewed approaches include confidence bands (also referred to as simultaneous confidence intervals) and confidence ellipsoids \cite{sachsConfidenceBandsSurvival2022, montieloleaSimultaneousConfidenceBands2019a, hooverAlgorithmsConfidenceCircles1984}, which are extensions of confidence intervals to parameter vectors. We provide a geometric perspective on how these different confidence regions relate and illustrate these ideas with three case studies. All illustrations were done using publicly available data and we provide code in different statistical software programs (SAS, R, Python). 

\section*{Data and Notation}

To orient our discussion, case studies with three types of simultaneous comparisons are considered: multiple causal effects, effect measure modification by a binary variable, and effect measure modification by a continuous variable. For each case study, we use data from the AIDS Clinical Trial Group (ACTG) 175, which was a randomized controlled trial of four different antiretroviral treatment combinations \cite{hammerTrialComparingNucleoside1996, zivichEstimatingMarginalStructural2025}. A total of 1571 observations were used. In all case studies, the action of interest was assignment to one of the following treatment regimens: two-drug (zidovudine-didanosine, zidovudine-zalcitabine) or one-drug (zidovudine). Baseline variables considered for modification were gender (male; female) and baseline CD4 count (cells per mm\textsuperscript{3}), a measure of immune function. The outcomes of interest were CD4 and CD8 cell counts at $20\pm5$ weeks.

We introduce some notation. Let $W_i$ denote the set of baseline variables, $A_i$ denote the action (i.e., assigned treatment), $Y_{i,j}$ indicate the outcome $j$ (i.e., $Y_{i,1}$: 20-week CD4 cell count, $Y_{i,2}$: 20-week CD8 cell count) for observation $i$. The observed data is then $O_i = (W_i, A_i, Y_{i,1}, Y_{i,2})$ for $i \in \{1,...,1571\}$. Potential outcomes are denoted using function notation (e.g., $Y_{i,j}(a)$). Further, let $E(\cdot)$ denote the expected value. Hereafter, $\alpha = 0.05$, which corresponds to two-sided 95\% confidence intervals, but the following discussion holds for any choice of $\alpha$ and one-sided confidence intervals.

\section*{Confidence Regions}

Here, we give a brief overview of the difference confidence regions: confidence \textit{intervals}, confidence \textit{bands}, and confidence \textit{ellipsoids}. The confidence interval is a range of meant to express the uncertainty of a \textit{single} parameter estimate \cite{sedgwickUnderstandingConfidenceIntervals2014}. For two-sided Wald-type confidence intervals, a critical value is selected based on the normal distribution justified by a central limit theorem. The choice of critical value if meant to ensure that the area under the normal distribution between plus or minus the critical value is equal to $1-\alpha$.

\begin{figure}[H]
	\centering
	\caption{Confidence regions for the average causal effect of two-drug versus one-drug antiretroviral therapy on 20-week CD4 and CD8 cell counts}
	\includegraphics[width=0.6\linewidth]{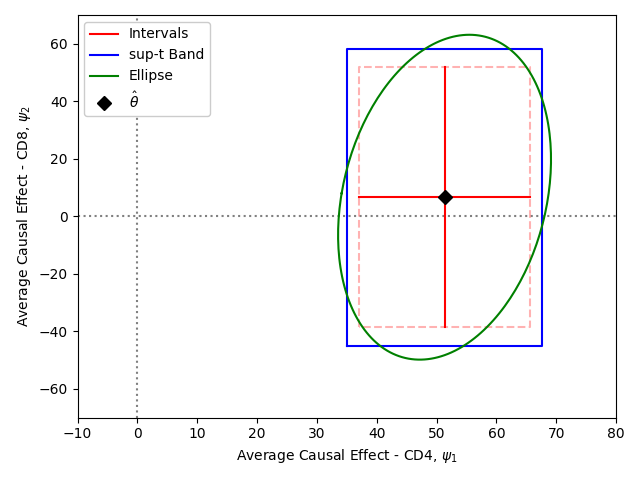}
	\floatfoot{The black diamond is the point estimate for the parameter vector. The red lines represent the confidence intervals, with the dashed lines denoting the inscribed rectangle corresponding to the pointwise confidence band. The blue rectangle corresponds to the sup-t bands and the green ellipse corresponds to the confidence ellipsoid.}
	\label{Figure1}
\end{figure}

Suppose we are interested in a \textit{collection}, or vector, of $k$ parameters. To contextualize the following introduction, let $k=2$. Here, we can think about a confidence region formed around the pair of point estimates (see Figure \ref{Figure1}). When thinking about coverage in this context, we consider whether that region contains the true point (i.e., the vector of the true parameter values). Confidence bands form a rectangle around the estimated parameter vector. Akin to how there are multiple ways to compute confidence intervals (e.g., Wald, score, likelihood), there are multiple ways to construct confidence bands \cite{sachsConfidenceBandsSurvival2022, montieloleaSimultaneousConfidenceBands2019a}. One simple way to compute the confidence bands is to take the intersection (or Cartesian product) of the confidence intervals to form a rectangle. The rectangle constructed from the intersection of the confidence bands is often referred to as a \textit{pointwise confidence band}. However, this rectangle formed by the confidence intervals is expected to contain the true point less than 95\% under repeated sampling. We can easily note this being the case since the coverage for each parameter in the vector is nominal, then the coverage for both parameters (i.e., two successes) will be below 95\%. This results from the area of our rectangle no longer being equal to $1-\alpha$ as in the one parameter case. The underlying coverage problem can also be thought of as akin to multiple testing, where each interval is constructed independent from the others. To remedy this deficiency, we review alternative ways to construct confidence bands. Confidence ellipsoids also aim to provide simultaneous coverage of parameter vectors but instead form an ellipse around the vector of point estimates. While both confidence bands and confidence ellipsoids have the same goal of simultaneous coverage, they have tradeoffs. To help further contextualize these ideas for the different confidence regions, we turn to the first case study.

\section*{Case Study 1: Multiple Effects}

To begin, suppose we are interested in the effect of two-drug treatment versus one-drug treatment on both CD4 and CD8 T cell counts at 20 weeks. This pair of parameters can be expressed using potential outcomes as
\begin{equation*}
	\psi = 
	\begin{bmatrix}
		\psi_1 \\
		\psi_2 
	\end{bmatrix}
	= 
	\begin{bmatrix}
		E[Y_1(1)] - E[Y_1(0)] \\
		E[Y_2(1)] - E[Y_2(0)]		
	\end{bmatrix} .
\end{equation*}
Given causal consistency (i.e., $Y_{i,j}(A_i) = Y_{i,j}$) \cite{coleConsistencyStatementCausal2009}, marginal exchangeability (i.e., $Y_j(a) \amalg A$ for $j \in \{1,2\}$) with positivity (i.e., $\Pr(A=a) > 0$ for $a \in \{0, 1\}$ in the population) \cite{hernanEstimatingCausalEffects2006, zivichPositivityIdentifiabilityEstimability2022}, it follows that 
\begin{equation*}
	\psi = 
	\begin{bmatrix}
		E[Y_1 \mid A = 1] - E[Y_1 \mid A = 0] \\
		E[Y_2 \mid A = 1] - E[Y_2 \mid A = 0]		
	\end{bmatrix} .
\end{equation*}
Therefore, both parameters are identified (i.e., expressed in terms of $O_i$). Note that marginal exchangeability and positivity are given by design of the trial. In addition to estimating $\psi$, we will also estimate the covariance matrix for the parameter vector, denoted by $V(\psi)$, using the empirical sandwich variance estimator \cite{rossMestimationCommonEpidemiological2024, mansourniaReflectionModernMethods2021}. Estimating $\psi$ with the data, we obtain point estimates of $\hat{\psi} = [\hat{\psi}_1, \; \hat{\psi}_2]^T = [51.3, \; 6.7]^T$, with $\hat{V}(\hat{\psi}_1) = 52.8$, $\hat{V}(\hat{\psi}_2) = 532.2$, and a covariance between the point estimates of $39.0$.

\subsection*{Confidence Regions}

Standard Wald-type two-sided confidence intervals can be computed via
\begin{equation*}
	{CI}_{\alpha} = 
	\begin{bmatrix}
		\hat{\psi}_1 \pm z_\alpha \sqrt{\hat{V}(\hat{\psi}_1)} \\ 
		\hat{\psi}_2 \pm z_\alpha \sqrt{\hat{V}(\hat{\psi}_2)}
	\end{bmatrix} ,
\end{equation*}
where $z_\alpha$ is the critical value (e.g., $z_\alpha \approx 1.96$ for $\alpha = 0.05$). The 95\% confidence intervals are then $[37.1, \; 65.6]$ for $\hat{\psi}_1$ and $[-38.6, \; 51.9]$ for $\hat{\psi}_2$. When reporting this pair of confidence intervals, we are implicitly led to interpret them together. We can visualize this as forming a rectangle around the pair of point estimates using the 95\% confidence intervals (i.e., pointwise confidence bands). While the coverage of each parameter independently will be 95\%, the rectangle will no longer cover the point corresponding to the true parameter vector at that level (i.e., simultaneous coverage).

To correct for the deficiency of the rectangle formed by the confidence intervals, we can consider other approaches to creating the confidence bands which enlarge the area of the previous rectangle. The general formula for these improved confidence bands is
\begin{equation*}
	{CB}_{\alpha} = 
	\begin{bmatrix}
		\hat{\psi}_1 \pm c_\alpha \sqrt{\hat{V}(\hat{\psi}_1)} \\ 
		\hat{\psi}_2 \pm c_\alpha \sqrt{\hat{V}(\hat{\psi}_2)}
	\end{bmatrix} ,
\end{equation*}
where $c_\alpha$ is a new, shared critical value replacing $z_\alpha$ that accounts for multiple comparisons shared between all estimates. These confidence bands are referred to as `1-parameter' confidence bands since there is a single shared $c_\alpha$ for all parameter estimates. There are various ways to compute $c_\alpha$. Perhaps the simplest in the Bonferroni method, where $\alpha$ is rescaled by the number of parameters (i.e., $c_\alpha = z_{\alpha/k}$). This rescaling produces a new set of wider univariate confidence intervals that are guaranteed to cover the true parameter vector at or above 95\% regardless of the covariance between parameters. Here, $c_\alpha = z_{\alpha/2} \approx 2.24$ and the corresponding confidence bands are $[35.0, \; 67.6]$ for $\hat{\psi}_1$ and $[-45.0, \; 58.4]$ for $\hat{\psi}_2$. While the Bonferroni correction does inflate the area, it inflates the area more than typically necessary as it does not account for the covariance between parameters (i.e., it assumes the estimates are independent). This excessive inflation can result in quite conservative (i.e., greater than 95\%) simultaneous coverage.

An alternative is the sup-t method (algorithm provided in Appendix 1), named as such since it is based on the largest (i.e., supremum) of the parameter-specific t-statistics \cite{montieloleaSimultaneousConfidenceBands2019a}. Like the Bonferroni method, the sup-t method widens the univariate confidence intervals but accounts for the estimated covariance between parameter estimators. In general, correlated estimates yield smaller values of $c_\alpha$, with the extreme case of perfectly correlated estimates yielding $c_\alpha = z_\alpha$. This adaptiveness can produce substantially narrower confidence bands with highly correlated parameters relative to the Bonferroni method but still provide correct simultaneous coverage unlike the rectangle formed by the confidence intervals. From a theoretical perspective, the sup-t confidence band has been shown to be the narrowest 1-parameter band that maintains nominal coverage \cite{montieloleaSimultaneousConfidenceBands2019a}. Here, $c_\alpha \approx2.23$ yielding sup-t confidence bands of $[35.1, \; 67.6]$ for $\hat{\psi}_1$ and $[-45.0, \; 58.3]$ for $\hat{\psi}_2$ (Figure \ref{Figure1}), which were not substantially different from the Bonferroni bands. If the covariance between parameters were larger, then the different between the Bonferroni and sup-t confidence bands would be expected to be larger.

Rather than adjust or modify the univariate confidence intervals, the confidence ellipsoid produces an ellipse around the point estimates. For the Wald confidence ellipsoid, that can be computed by rescaling and shifting the unit circle (rescaling determined by the eigenvalue and eigenvector of the covariance matrix, shift determined by the parameter vector). Algorithms for computing confidence ellipsoids are given in more detail elsewhere \cite[p.~811-817]{pressNumericalRecipes3rd2007}\cite{hooverAlgorithmsConfidenceCircles1984}. To report a confidence ellipsoid in an easily interpretable way, a graphical presentation is used (Figure \ref{Figure1}). Unlike the previous confidence regions, the confidence ellipsoid can no longer be directly summarized with pairs of values.

While both confidence bands and confidence ellipsoids are meant to simultaneous inference, important differences can be gleaned from their visual differences in Figure \ref{Figure1}. Most obvious is the difference between their geometric shapes. Further, the area covered by the ellipse is smaller than the sup-t rectangle. So, some parameter combinations are only covered by one of the confidence regions. Therefore, the inference being made between these two confidence regions differs, despite both producing simultaneous coverage. In particular, the Wald confidence ellipse connects back to a joint Wald test of the null hypothesis (all parameters are equal to zero) whereas the confidence bands correspond to marginal tests of each null (with control for multiple comparisons when not based on confidence intervals). While confidence ellipsoids produce more precise inference (in terms of a smaller area), confidence ellipsoids can be more difficult to present and interpret than confidence bands.

\section*{Case Study 2: Effect Modification -- Binary}

Now suppose interest was in whether the effect of two-drug treatment versus one-drug treatment on 20-week CD4 was modified by gender on the additive scale. With a trial, we can address this question by estimating the parameters of the following linear model
\begin{equation*}
	E[Y_1(a) \mid V; \beta] = \beta_0 + \beta_1 a + \beta_2 V + \beta_3 aV
\end{equation*}
where $V$ denotes gender and $\beta = [\beta_0, \beta_1, \beta_2, \beta_3]$ \cite{zivichEstimatingMarginalStructural2025, chibaMarginalStructuralModels2009}. We assume causal consistency and exchangeability with positivity by $V$ (which is still given by design). Here, linear regression is used to estimate $\beta$ (Table \ref{Table1}). 

\begin{table}[H]
	\centering
	\caption{Estimated parameters and corresponding confidence regions for the model describing the departure from additivity of antiretroviral therapy on CD4 cell counts by gender}
	\begin{tabular}{llcccc}
		\hline
		&                         & $\beta_0$          & $\beta_1$         & $\beta_2$        & $\beta_3$         \\ \cline{3-6} 
		\multicolumn{2}{l}{Estimate}      & 356.8              & -25.5             & 41.4             & 12.6              \\
		\multicolumn{2}{l}{Confidence}    &                    &                   &                  &                   \\
		& Intervals               & {[}328.8, 384.8{]} & {[}-55.9, 5.0{]}  & {[}5.4, 77.4{]}  & {[}-26.6, 51.7{]} \\
		& Band -- Bonferroni      & {[}321.1, 392.5{]} & {[}-64.3, 13.4{]} & {[}-4.5, 87.2{]} & {[}-37.4, 62.5{]} \\
		& Band -- sup-t           & {[}324.1, 389.6{]} & {[}-61.1, 10.2{]} & {[}-0.7, 83.5{]} & {[}-33.3, 58.4{]} \\
		\multicolumn{2}{l}{Region widths} &                    &                   &                  &                   \\
		& Intervals               & 56.0               & 60.9              & 72.0             & 78.3              \\
		& Band -- Bonferroni      & 71.4               & 77.7              & 91.8             & 99.8              \\
		& Band -- sup-t           & 65.5               & 71.3              & 84.2             & 91.6              \\ \hline
	\end{tabular}
	\label{Table1}
\end{table}

\subsection*{Confidence Regions}

There are two ways the results from this model might be reported. First, one might only report the estimate for $\beta_3$. In this case, reporting the standard confidence intervals for $\hat{\beta}_3$ is appropriate. Another option is to present the stratum specific effects (i.e., $\beta_1$ and $\beta_1 + \beta_3$). Finally, the full set of coefficients could be provided, as done in Table 1. In either of the latter cases, the parameter of interest is better described as a vector and thus better summarized using confidence regions with simultaneous coverage.

As before, confidence intervals for each $\beta$ can be reported (Table \ref{Table1}). However, when thinking about these confidence intervals simultaneously covering $\beta$, one is led astray. Confidence bands are again simple to compute and present here. Table \ref{Table1} provides both Bonferroni and sup-t confidence bands. Here, a more noticeable difference is observed between the Bonferroni and sup-t bands when comparing the width of the intervals (upper limit minus the lower limit). When the hypervolume (4-dimensional volume) is computed, the Bonferroni bands produce a hyperrectangle with a hypervolume that is 1.4 times larger than that of the sup-t bands.

Here, we do not report the confidence ellipsoid, as the ellipsoid cannot simply be reported in a table and plotting a 4D ellipsoid is also challenging. One solution is to instead plot all the pairwise ellipsoids (a plot with 6 panels here) \cite{friendlyHEPlotsMultivariate2007}. However, this solution becomes untenable as the size of the parameter vector becomes larger. The use of confidence ellipsoids in these settings also represents a large shift in how epidemiologists report and present results. Given these limitations of confidence ellipsoids, confidence bands seem to be preferred over ellipsoids when there are more than two parameters of interest.

\section*{Case Study 3: Effect Modification -- Continuous}

For the final example, suppose interest was in whether the effect of two-drug treatment versus one-drug treatment on 20-week CD4 by varying baseline CD4 levels on the additive scale. We can address this question with the following linear model
\begin{equation*}
	E[Y_1(a) \mid X; \gamma] = \gamma_0 + \gamma_1 a + s(X) \gamma_s^T + s(X) a \gamma_{a,s}^T
\end{equation*}
where $X$ denotes baseline CD4, $s(\cdot)$ is a function that creates a spline design matrix for the input variable, $\gamma_s$ and $\gamma_{a,s}$ are both vectors and $\gamma = [\gamma_0, \gamma_1, \gamma_s, \gamma_{a,s}]$ \cite{howeSplinesTrendAnalysis2011a, discacciatiEstimatingPresentingNonlinear2025}. Again, because $A$ was marginally randomized, linear regression can be used to estimate the parameters of this model. While the parameter estimates can be reported in a table, it is generally more interpretable to plot how effects vary by $X$ with smoothing (e.g., splines) \cite{discacciatiEstimatingPresentingNonlinear2025}. Here, we report the conditional average causal effect, which is defined as $\gamma_1 + s(x) \gamma_{a,s}^T$ for a given value $x$, using a plot.

\subsection*{Confidence Regions}

With $\hat{\gamma}$ and $\hat{V}(\hat{\gamma})$ the line corresponding to the conditional average causal effect and the associated 95\% confidence intervals can be computed. Each 95\% confidence interval then claims to cover the parameter at its value of $x$ (i.e., $E[Y_1(1) - Y_1(0) \mid X=x]$) at the 0.95 level. However, confidence intervals are often presented as a shaded region encompassing the point estimates. Implicitly, the corresponding shaded region are conveying a confidence region for the conditional average causal effect \textit{function}, not at a specific point. Since functions can be thought of as infinite-dimension vectors, confidence regions that provide simultaneous coverage are more appropriate.

\begin{figure}[H]
	\centering
	\caption{Confidence regions for the conditional average causal effect of two-drug versus one-drug antiretroviral therapy on 20-week CD4 cell counts by baseline CD4.}
	\includegraphics[width=0.8\linewidth]{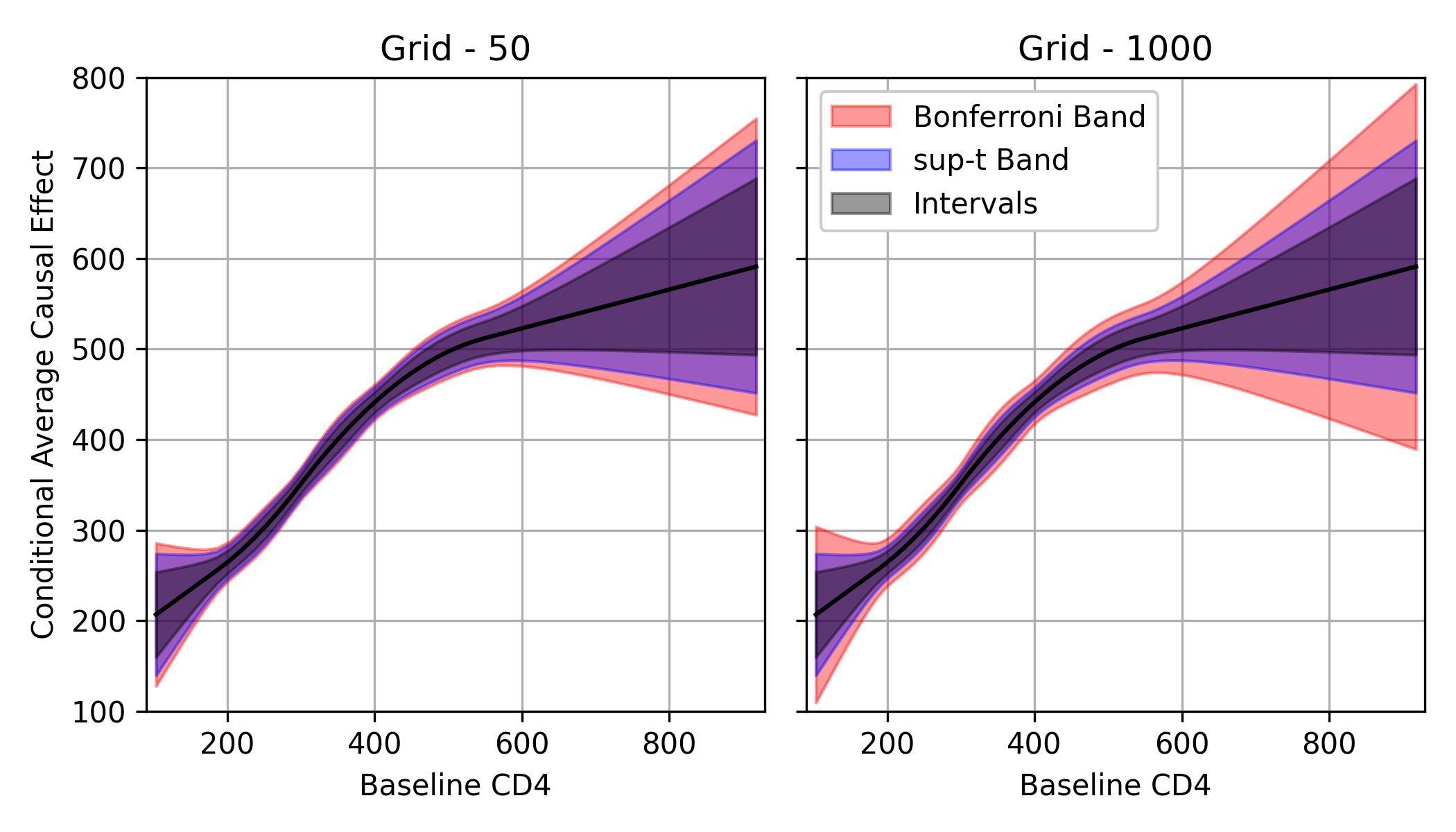}
	\floatfoot{The left plot is based on a grid of 50 evenly spaced values of baseline CD4 to predict for and the right plot is based on a grid of 1000 evenly spaced values of baseline CD4 to predict for.  The black line represents the point estimate of the conditional average causal effect by baseline CD4. The gray shaded region corresponds to confidence intervals, or pointwise confidence band. The red shaded region depicts the Bonferroni confidence bands. The blue shaded region, which appears purple since it is nested within the Bonferroni bands, are the sup-t confidence band.}
	\label{Figure2}
\end{figure}

When plotting the estimate of the functions, we evaluate the conditional average causal effect across a grid of evenly spaced values with the range of observed baseline CD4 counts. These point estimates are then connected by lines. Generally, the number of points in the grid is arbitrary, with the goal of approximating the continuous function. Here, we consider two grids with varying numbers of equally spaced points: 50 and 1000. For the confidence bands, both Bonferroni and sup-t confidence bands can be computed. When computing Bonferroni confidence bands, the critical value depends on the number of points in our grid. Therefore, the critical value is 3.29 for the grid of 50 points and 4.06 for the grid with 1000 points. As shown in Figure \ref{Figure2}, the choice of grid widths can produce a noticeable change in the Bonferroni confidence bands. Given that the grid is an arbitrary choice for visualization purposes, the Bonferroni bands being sensitive to this choice highlights another one of their weaknesses. From a statistical theory perspective, the Bonferroni method does not generalize to infinite-dimensions. However, the sup-t confidence bands are defined for functions. Further, the critical value from the sup-t method is stable with a changing grid size due to the sup-t method incorporating the covariance between parameters in the vector. As the grid width shrinks, the correlation between nearby parameters increases. As the correlation is incorporated into the sup-t algorithm, the sensitivity to the grid choice is effectively removed. Here, the sup-t critical value is approximately 2.80. As in the previous case study, presenting confidence ellipsoids is difficult. Here, the difficulties are compounded by the choice of the grid and then needing to visualize each pair of grid points (or at least those nearby). Therefore, sup-t confidence bands are recommended for displaying confidence regions around functions.

\section*{Discussion}

Here, we provided an introduction to confidence intervals, bands, and ellipsoids. To summarize our key message, the inferential approach an epidemiologist leverages should incorporate whether interest is in a single parameter or a parameter vector (e.g., multiple regimes or outcomes, joint effects, effect measure modification, dose-response, functions). Confidence bands are preferred over ellipsoids due to their similarity to confidence intervals in presentation and preferred over intervals due to their valid statistical coverage properties for parameter vectors. For computing confidence bands, the sup-t method is preferred due to its desirable statistical properties and the ease with which it can be computed given a covariance matrix. These statistical properties and simulations illustrating them have been reported elsewhere in more detail \cite{ashConfidenceBandsHypothesis2022,montieloleaSimultaneousConfidenceBands2019a}. To ease adoption, we provide example code in SAS, R, and Python but tehre is also open-source software for computing confidence bands \cite{hothornMultcompSimultaneousInference2025, greiferFwbFractionalWeighted2025}.

Reporting confidence bands is in line with the perspective shift in epidemiology from null hypothesis significance testing to focusing on the values of parameter estimates \cite{gardnerConfidenceIntervalsRather1986, stangStatisticalInferenceAbstracts2017b}. An ongoing debate is whether the qualifier `confidence' should be used for `confidence intervals', with `uncertainty' and `compatible' offered as alternatives \cite{amrheinScientistsRiseStatistical2019, gelmanAreConfidenceIntervals2019, greenlandCurbResearchMisreporting2022, rafiSemanticCognitiveTools2020}. Neither term has become common in the epidemiology literature and thus we retained the standard terminology of `confidence'. Regardless, as the field of epidemiology considers such a terminological shift, the names of the described intervals, bands, and ellipsoids are easily modified and the key messages of this article remain relevant. Another debate is the interpretation of probabilities. The confidence regions described in this paper are under a frequentist (or sampling theory) perspective, which appears to be the implicit framework of many epidemiology studies. Others have argued that a Bayesian perspective is better able to resolve multiple-comparison questions based on the criticism that it is unclear which `hypotheses' should be corrected for and it can seem arbitrary from a frequentist perspective \cite{sjolanderFrequentistBayesianApproaches2019, berryBayesianPerspectivesMultiple1999}. By focusing on which parameters correspond to the scientific question of interest, we find that is becomes clear which comparisons to correct for under a frequestist perspective \cite{greenlandMultipleComparisonsControversies2019}. Regardless, corresponding Bayesian sup-t credible bands can also be computed \cite{montieloleaSimultaneousConfidenceBands2019a}. Alternatively, one could use hierarchical, or multi-level, models under either a Bayesian or frequentist paradigm to correct for multiple comparisons \cite{sjolanderFrequentistBayesianApproaches2019, greenlandMultipleComparisonsControversies2019, greenlandAnalysisGoalsErrorcost2021}. Such hierarchical models may produce more precise inference by exploiting the correlation between parameters for point estimation, unlike the approach reviewed here.

The case studies examined here were in the context of a randomized trial. Our same message also applies to observational research. Within observational studies, additional parameters need to be estimated but are not of interest (i.e., nuisance parameters). Published papers using confidence bands with observational methods include the following references \cite{brumbackSensitivityAnalysesUnmeasured2004, plattInformationCriterionMarginal2013, kovalchikSearchLostTime2020}. As an example, the propensity score in inverse probability weighting estimators is often estimated using logistic regression \cite{joffeInvitedCommentaryPropensity1999, robinsMarginalStructuralModels2000}. In these settings, the corrected critical value should only account for the parameters of interest \cite{greenlandAnalysisGoalsErrorcost2021}, so only the covariance matrix for the interest parameters is needed to construct the sup-t confidence bands. However, the uncertainty in the estimated nuisance parameters should be incorporated into the estimate covariance matrix. Estimation of the covariance matrix that does appropriately incorporate the uncertainty of nuisance parameters can easily be accomplished with estimating equations and the empirical sandwich variance estimator \cite{rossMestimationCommonEpidemiological2024, stefanskiCalculusMEstimation2002}. As the outer produce of the influence functions is equivalent to the empirical sandwich variance estimator \cite{coleFiveFactsInfluence2025}, influence functions can be used in a similar way. In Appendix 2, we replicate the three case studies with an inverse probability weighting estimator adjusting for baseline covariates with estimating equations. When the covariance matrix cannot be easily estimated, sup-t confidence bands can also be computed for `regular' parameters using a bootstrap procedure \cite{montieloleaSimultaneousConfidenceBands2019a, greiferFwbFractionalWeighted2025}. One final caveat is that the described confidence regions do not account for variable selection procedures occurring outside the estimation process (e.g., ad hoc model choice by Akaike information criterion or likelihood ratio test).  Additional care is needed for estimation of the covariance matrix, and thus the confidence bands, when data adaptive procedures are used. To summarize these implications, the validity of the sup-t confidence bands depends on valid estimation of the parameters of interest and their covariance (or the validity of the bootstrap procedure) but is separate from that estimation process.

Given epidemiology's transition from null hypothesis significance testing to focusing on parameter estimation, our methods of inference should be similarly focused. When the scientific question is addressed through estimation of a single parameter, then confidence intervals are appropriate. When interest is in a collection of parameters, confidence bands, which extend inference from a parameter to a vector, are more appropriate to report.

\section*{Acknowledgments}

Research reported in this publication was supported by National Institute of Allergy and Infectious Diseases of the National Institutes of Health under award numbers K01AI177102 (PNZ), R01AI157758 (PNZ, SRC, JKE), P30AI50410 (PNZ, SRC), K99MH133985 / R00MH133985 (LMM). The content is solely the responsibility of the authors and does not necessarily represent the official views of the National Institutes of Health. \\
Data and computing code availability: Data and code to implement the confidence regions in SAS, R, and Python is available at \url{https://github.com/pzivich/publications-code}

\small
\bibliography{references}{}
\bibliographystyle{ieeetr}

\newpage
\normalsize

\section*{Appendix}

\setcounter{figure}{0}
\renewcommand{\thefigure}{A\arabic{figure}}
\setcounter{table}{0}
\renewcommand{\thetable}{A\arabic{table}}

\subsection*{Appendix 1: sup-t Confidence Bands Algorithm}

Let $\psi = (\psi_1, ..., \psi_p)$ denote the parameter vector of interest, $V(\psi)$ denote the covariance matrix for the parameter vector of interest, and hats denote estimates. The sup-t critical value can be estimated using the following algorithm
\begin{enumerate}
	\item Compute the point estimates, $\hat{\psi}$, and the covariance estimates, $\hat{V}(\hat{\psi})$, for the parameter vector of interest.
	\item For each $j$ from $1$ to $m$
	\begin{itemize}
		\item[a.] Draw a random vector, $\delta_j$, from a multivariate normal distribution centered at zero with a covariance matrix of $\hat{V}(\hat{\psi})$.
		\item[b.] Compute $\delta_j'$ by dividing $\delta_j$ by the estimated standard error, $\sqrt{\text{diag}\left(\hat{V}(\hat{\psi})\right)}$ where $\text{diag}(\cdot)$ denotes taking the diagonal of the covariance matrix (i.e., the variance).
		\item[c.] Take the absolute value of $\delta_j'$ and find the maximum. Store this maximum, denoted as $\Delta_j$
	\end{itemize}
	\item From the $m$ different maximum values, $[\Delta_1, ..., \Delta_m]$, take the $(1 - \alpha) \times 100$ percentile as the estimate of $c_\alpha$, indicated by $\hat{c}_\alpha$.
\end{enumerate}
Here, $m$ should be large (e.g., 10,000) to minimize the estimation error of $c_\alpha$. The confidence bands can be computed by replacing $c_\alpha$ with $\hat{c}_\alpha$ in the standard formula for the confidence interval. Note that this algorithm requires that the estimated covariance matrix for the interest parameters $\hat{V}(\hat{\psi})$ is available. There are also alternative algorithms available to compute the sup-t $c_\alpha$ \cite{genzMvtnormMultivariateNormal2025}. When the estimated covariance matrix $\hat{V}(\hat{\psi})$ is not available, a bootstrapping procedure can be used \cite{montieloleaSimultaneousConfidenceBands2019a}.

\subsection*{Appendix 2: Confidence Bands with Inverse Probability Weighting}

To illustrate the use of confidence bands with observational research methods, we replicate the main examples from the paper but now treat the ACTG 175 as if it were can observational study. Here, the following baseline variables ($W$) are treated as confounding variables: age (years), race (white; black), history of injection drug use (yes; no), Karnofsky score category (70-89; 90-99; 100), baseline CD4, and baseline CD8. While not necessary, applying methods that adjust for baseline covariates, like inverse probability weighting (IPW), to trials can meaningfully increase precision \cite{morrisPlanningMethodCovariate2022}. In the following sections, we assume causal consistency \cite{coleConsistencyStatementCausal2009}, conditional exchangeability given $W$ for both potential outcomes (i.e., $Y_j(a) \amalg A \mid W$ for $j \in \{1,2\}$) with positivity (i.e., $\Pr(A=a \mid W=w) > 0$ for $a\in\{0,1\}$ and all unique values of $w$ in the population) \cite{hernanEstimatingCausalEffects2006, zivichPositivityIdentifiabilityEstimability2022}. Therefore, both parameters are identified (i.e., expressable in terms of $O_i$). Given that ACTG 175 was a trial, the exchangeability with positivity assumption is met by-design.

To estimate the interest parameters, we use an IPW estimator \cite{robinsMarginalStructuralModels2000, breskinExploringSubtletiesInverse2018, zivichEstimatingMarginalStructural2025}. Hereafter, $\Pr(A=a \mid W)$, the propensity score \cite{joffeInvitedCommentaryPropensity1999}, is estimated using a logistic regression model. To simplify consistent covariance estimation, we frame the following IPW estimators as a set of estimating equations \cite{rossMestimationCommonEpidemiological2024, stefanskiCalculusMEstimation2002}, and estimate the covariance of the parameters using the empirical sandwich variance estimator. An estimating equation is defined as $E[g(O_i; \theta)] = 0$, where $g$ is the estimating function and $\theta$ is the parameter vector. Let $V(\theta)$ denote the covariance matrix for the parameters. Importantly for the application here, the empirical sandwich variance estimator allows consistent estimation of the covariance matrix which will be used with the sup-t confidence bands.

\subsubsection*{Case Study 1}

Given causal consistency and exchangeability with positivity, it follows that
\begin{equation*}
	\psi = 
	\begin{bmatrix}
		E\left( \frac{Y_1 A}{\Pr(A=1 \mid W)} \right) - E\left( \frac{Y_1 (1-A)}{\Pr(A=0 \mid W)} \right) \\
		E\left( \frac{Y_2 A}{\Pr(A=1 \mid W)} \right) - E\left( \frac{Y_2 (1-A)}{\Pr(A=0 \mid W)} \right) \\
	\end{bmatrix}
\end{equation*}
and the estimating function corresponding to the IPW estimator is
\begin{equation*}
	g(O_i; \theta) = 
	\begin{bmatrix}
		(A_i - \hat{A}_i) \mathbb{W}_i^T \\
		\frac{A_i}{\hat{A}_i} (Y_{i,1} - \mu_1) \\
		\frac{1-A_i}{1-\hat{A}_i} (Y_{i,1} - \mu_0) \\
		\frac{A_i}{\hat{A}_i} (Y_{i,2} - \omega_1) \\
		\frac{1-A_i}{1-\hat{A}_i} (Y_{i,2} - \omega_0) \\
		(\mu_1 - \mu_0) - \psi_1 \\
		(\omega_1 - \omega_0) - \psi_2 \\
	\end{bmatrix}
\end{equation*}
where $\hat{A}_i = \text{expit}(\mathbb{W}_i \eta^T)$, $\mathbb{W}_i$ is the $i$ row of the design matrix build from $W$, $\eta$ is the parameter vector for the propensity score model, $\text{expit}(x) = \left\{ 1 + \exp(-x) \right\}^{-1}$, and $\theta = (\eta, \mu_1, \mu_0, \omega_1, \omega_0, \psi_1, \psi_2)$. Here, the first element of the vector is the logistic model for the propensity scores. The second and third are the Hajek IPW estimators for CD4 under two-drug and one-drug antiretroviral therapy, respectively. The fourth and fifth are the Hajek IPW estimators for CD8 under two-drug and one-drug antiretroviral therapy, respectively. The sixth is the average causal effect for CD4 and the seventh is the average causal effect for CD8.

\begin{figure}[H]
	\centering
	\caption{Confidence regions for the average causal effect of two-drug versus one-drug antiretroviral therapy on 20-week CD4 and CD8 cell counts using inverse probability weights}
	\includegraphics[width=0.6\linewidth]{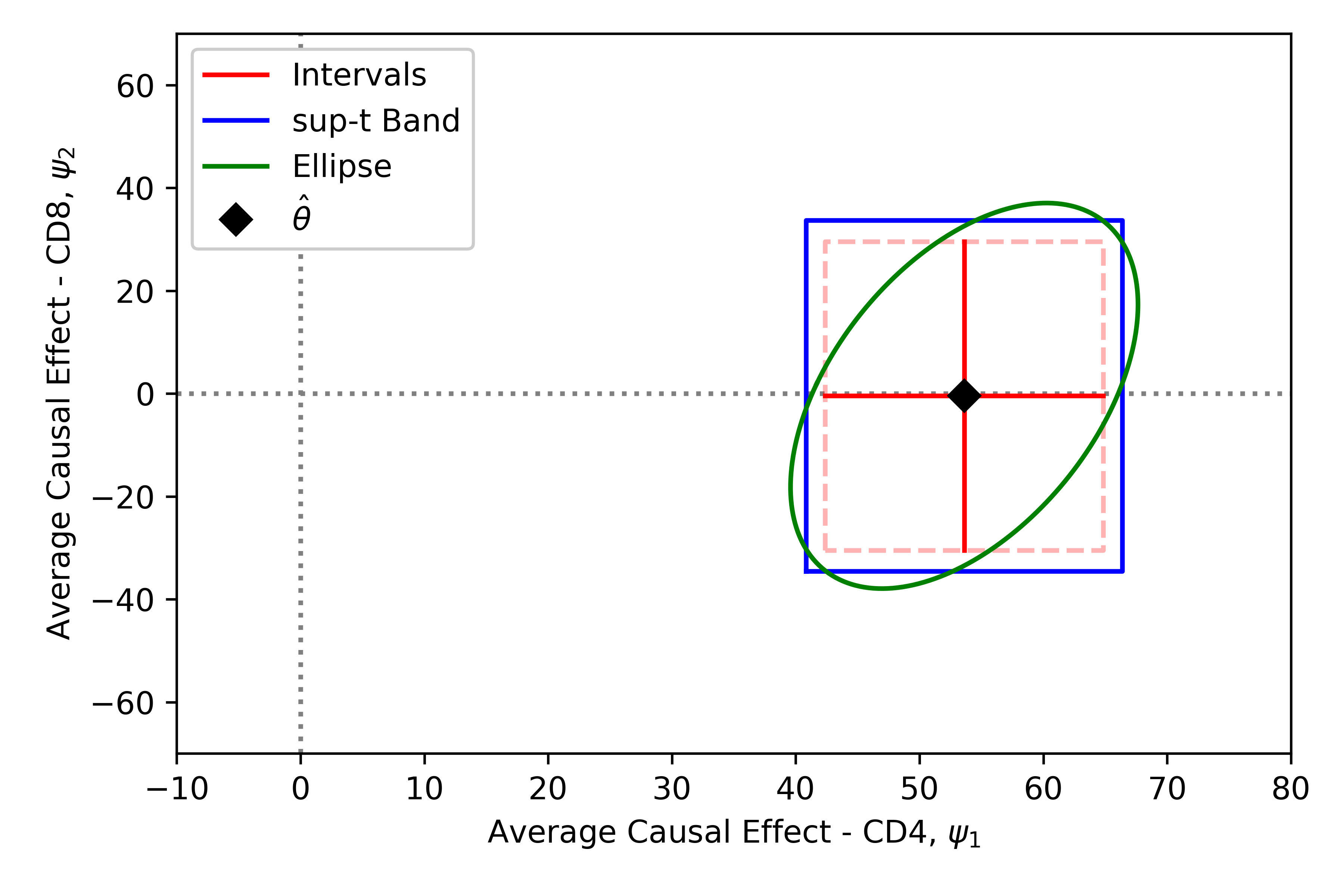}
	\floatfoot{The black diamond is the point estimate for the parameter vector. The red lines represent the confidence intervals, with the dashed lines denoting the inscribed rectangle corresponding to the pointwise confidence band. The blue rectangle corresponds to the sup-t bands and the green ellipse corresponds to the confidence ellipsoid.}
	\label{FigureA1}
\end{figure}

The estimated effects were $\hat{\psi}_1 = 53.6$ and  $\hat{\psi}_2 = -0.4$. The 95\% confidence intervals were $(42.4, \; 64.8)$ and $(-30.5, \; 29.6)$, respectively. The Bonferroni critical value was 2.24 with confidence bands of $(40.8, \; 66.5)$ and $(-34.8, \; 33.9)$, respectively. For the sup-t method, $c_\alpha$ is estimated using the same algorithm as before. However, only the parameter vector $[\hat{\psi}_1, \hat{\psi}_2]$ and covariance matrix for $\hat{\psi}_1$ and $\hat{\psi}_2$ is used. Note that the empirical sandwich variance estimator incorporates the uncertainty of the propensity score model (i.e., the nuisance model) into the estimation of $V(\theta)$, but the sup-t confidence bands only need to consider those elements that correspond to the parameters of interest (i.e., the nuisance parameters are ignored for $c_\alpha$). The sup-t critical value was 2.23 with confidence bands of $(40.8, \; 66.4)$ and $(-34.5, \; 33.7)$, respectively. Similarly, the confidence ellipsoid only considers $\psi$. Confidence regions are visualized in Figure \ref{FigureA1}. As can be seen when contrasting with Figure \ref{Figure1}, use of the IPW estimator noticeably increased precision, most notably for the average causal effect on CD8.

\subsubsection*{Case Study 2}

Now consider estimating the model as described in the main paper but with weighted linear regression to account for confounding. The estimating function for the corresponding IPW marginal structural model is
\begin{equation*}
	g(O_i; \theta) = 
	\begin{bmatrix}
		(A_i - \hat{A}_i) \mathbb{W}_i^T \\
		\left( \frac{A_i}{\hat{A}_i} + \frac{1-A_i}{1-\hat{A}_i} \right) \left\{ Y_{i,1} - \hat{Y}_{i,1} \right\} \mathbb{X_i}^T
	\end{bmatrix}
\end{equation*}
where $\hat{A}_i$ and $\mathbb{W}_i$ are defined the same as the previous estimating function. Here $\mathbb{X}_i$ is hte design matrix for the marginal structural model (i.e., $\mathbb{X}_i = (1, A_i, V_i, A_i V_i)$) and $\hat{Y}_{i,1} = \mathbb{X}_i \beta^T$. Again, the confidence bands only need to consider the parameters in $\beta$. Results are reported in Appendix Table \ref{TableA1}

\begin{table}[H]
	\centering
	\caption{Estimated parameters and corresponding confidence regions for the marginal structural model describing effect measure modification of antiretroviral therapy by gender estimated using inverse probability weights}
	\begin{tabular}{llcccc}
		\hline
		&                         & $\beta_0$          & $\beta_1$         & $\beta_2$        & $\beta_3$         \\ \cline{3-6} 
		\multicolumn{2}{l}{Estimate}      & 356.0              & -26.4             & 42.4             & 13.6              \\
		\multicolumn{2}{l}{Confidence}    &                    &                   &                  &                   \\
		& Intervals               & {[}327.6, 384.3{]} & {[}-58.1, 5.3{]}  & {[}7.0, 77.7{]}  & {[}-26.4, 53.6{]} \\
		& Band -- Bonferroni      & {[}319.1, 392.1{]} & {[}-66.8, 14.0{]} & {[}-2.7, 87.4{]} & {[}-37.4, 64.6{]} \\
		& Band -- sup-t           & {[}323.2, 388.7{]} & {[}-63.1, 10.3{]} & {[}1.5, 83.3{]} & {[}-32.7, 59.9{]} \\
		\multicolumn{2}{l}{Region widths} &                    &                   &                  &                   \\
		& Intervals               & 56.7               & 63.4              & 70.7             & 80.0              \\
		& Band -- Bonferroni      & 72.2               & 80.7              & 90.1             & 102.0              \\
		& Band -- sup-t           & 65.6               & 73.3              & 81.8             & 92.6              \\ \hline
	\end{tabular}
	\label{TableA1}
\end{table}

\subsubsection*{Case Study 3}

\begin{figure}[H]
	\centering
	\caption{Confidence regions for the conditional average causal effect of two-drug versus one-drug antiretroviral therapy on 20-week CD4 cell counts by baseline CD4 using inverse probability weights.}
	\includegraphics[width=0.8\linewidth]{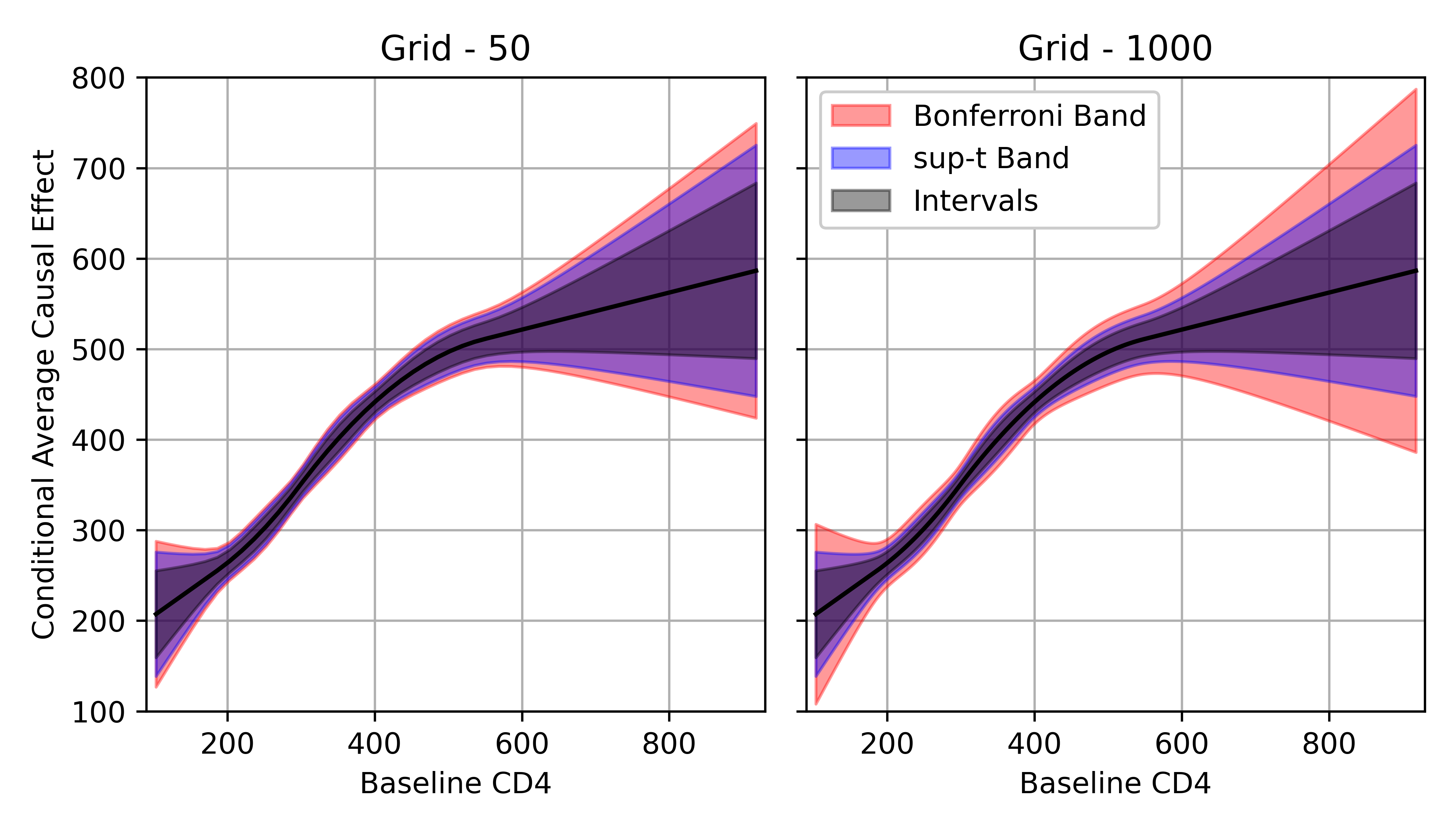}
	\floatfoot{The left plot is based on a grid of 50 evenly spaced values of baseline CD4 to predict for and the right plot is based on a grid of 1000 evenly spaced values of baseline CD4 to predict for.  The black line represents the point estimate of the conditional average causal effect by baseline CD4. The gray shaded region corresponds to confidence intervals, or pointwise confidence band. The red shaded region depicts the Bonferroni confidence bands. The blue shaded region, which appears purple since it is nested within the Bonferroni bands, are the sup-t confidence band.}
	\label{FigureA2}
\end{figure}

Again, consider estimating the model as described in the main paper but with weighted linear regression to account for confounding. The estimating function for the IPW marginal structural model is the same as in Case Study 2 except for the design matrix of the marginal structural model. For the grid of $m$ baseline CD4 values, $[x_1, ..., x_m]$, a corresponding design matrix can be generated, denoted by $\mathcal{X}$. For computing the confidence bands, let $\hat{y}$ denote the predicted conditional average causal effect from the estimate marginal structural model parameters. The covariance for all $\hat{y}$ can be computed from the estimated covariance matrix of the marginal structural model parameters, $\hat{V}(\hat{\gamma})$, via $\hat{V}(\hat{\gamma}) \cdot \mathcal{X} \cdot \hat{V}(\hat{\gamma})^T$. From this covariance matrix and $\hat{y}$, the confidence intervals and bands can be computed. Figure \ref{FigureA2} provides the corresponding plots for $m = 50$ and $m=1000$.

\end{document}